\documentclass[10pt,letterpaper]{article}
\usepackage[top=0.85in,left=2.75in,footskip=0.75in,marginparwidth=2in]{geometry}

\usepackage[utf8]{inputenc}

\usepackage{cite}
\usepackage{siunitx,amsmath}
\usepackage{nameref,hyperref, cleveref, bm}

\usepackage[right]{lineno}

\usepackage{microtype}
\DisableLigatures[f]{encoding = *, family = * }

\raggedright
\usepackage{changepage}

\usepackage[aboveskip=1pt,labelfont=bf,labelsep=period,singlelinecheck=off]{caption}

\makeatletter
\renewcommand{\@biblabel}[1]{\quad#1.}
\makeatother

\usepackage{lastpage,fancyhdr,graphicx}
\usepackage{epstopdf}
\fancyheadoffset[L]{2.25in}
\fancyfootoffset[L]{2.25in}

\usepackage{color}

\definecolor{Gray}{gray}{.25}

\usepackage{graphicx}

\usepackage{sidecap}

\usepackage{wrapfig,romannum, url}
\usepackage[pscoord]{eso-pic}
\usepackage[fulladjust]{marginnote}
\reversemarginpar

\begin{document}
	\vspace*{0.35in}
	\pagenumbering{arabic}
	\begin{flushleft}
		{\Large
			\textbf\newline{Eigen Microstates and Their Evolution of Global Ozone at Different Geopotential Heights}
		}
		\newline
		\\
		Xiaojie Chen\textsuperscript{1},
		Na Ying\textsuperscript{2},
		Dean Chen\textsuperscript{3},
		Yongwen Zhang\textsuperscript{4},
		Bo Lu\textsuperscript{5},
		Jingfang Fan\textsuperscript{1,(a)},
		Xiaosong Chen\textsuperscript{1,(b)}
		\\
		\bigskip
		\bf{1} School of Systems Science, Beijing Normal University, Beijing, China, 100875
		\\
		\bf{2} China State Key Laboratory of Environmental Criteria and Risk Assessment, Chinese Research Academy of Environmental
		Sciences, Beijing, China, 100012
		\\
		\bf{3} Institute for Atmospheric and Earth System Research/Physics, Faculty of Science, University of Helsinki, P.O. Box 68, 00014 Helsinki, Finland
		\\
		\bf{4} Data Science Research Center, Faculty of Science, Kunming University of Science and Technology, Kunming, China, 650500
		\\
		\bf{5} Laboratory for Climate Studies and CMA-NJU Joint Laboratory for Climate Prediction Studies, National Climate Center, China Meteorological Administration, Beijing, China, 100081\\
		\bigskip
		(a) jingfang@bnu.edu.cn,  (b) chenxs@bnu.edu.cn
		
	\end{flushleft}
	
	\section*{Abstract}
	Studies on stratospheric ozone have attracted much attention due to its serious impacts on climate changes and its important role as a tracer of Earth's global circulation.
	Tropospheric ozone as a main atmospheric pollutant damages human health as well as the growth of vegetation.
	Yet there is still a lack of a theoretical framework to fully describe the variation of ozone.
	To understand ozone's spatiotemporal variance, we introduce the eigen microstate method to analyze the global ozone mass mixing ratio (OMMR) between 1979-01-01 and 2020-06-30 at 37 pressure layers.
	We find that eigen microstates at different geopotential heights can capture different climate phenomena and modes.
	Without deseasonalization, the first eigen microstates capture the seasonal effect, and reveal that the phase of the intra-annual cycle moves with the geopotential heights.
	After deseasonalization, by contrast, the collective patterns from the overall trend, ENSO, QBO, and tropopause pressure are identified by the first few significant eigen microstates.
	The theoretical framework proposed here can also be applied to other complex Earth systems.
	
	
	
	\section{Introduction}
	
	Ozone plays different roles in life on Earth: the stratosphere ozone absorbs most ultraviolet radiation protecting lives; but the ground-level ozone is a secondary pollutant, exerting a severe threat to plant and public health \cite{brauerAmbientAirPollution2016, karimiConcentrationsHealthEffects2019, xuEstimatingImpactGround2021}.
	It has been reported that the stratosphere ozone is decreasing year by year \cite{wmoworldmeteorologicalorganizationScientificAssessmentOzone2018} while the ground-level ozone has a rising trend \cite{fengGroundlevelO3Pollution2015}, resulting in a serious challenge to human society.
	The variation of ozone is controlled by chemical and physical processes, which are affected by both natural processes and anthropogenic activities \cite{staehelinOzoneTrendsReview2001}.
	Due to the nonlinear feedbacks, multiple interactions and complex structures of the atmospheric system, the understanding of the the spatiotemporal variance of ozone is challenging.
	
	The impacts of atmosphere circulations on ozone have been widely studied \cite{nowackLargeOzonecirculationFeedback2015,polvaniStratosphericOzoneDepletion2011,jockelAtmosphericChemistryGeneral2006,bronnimannVariabilityTotalOzone2000,oltmansSurfaceOzoneDistributions1986}, mostly focusing on the upper troposphere and lower stratosphere regions.
	Herman \textit{et al.} \cite{hermanGlobalAverageOzone1991} deduced the ozone depletion trend, quasi-biennial oscillation (QBO) and a possible 11-year solar cycle variation from globally averaged TOMS data.
	Later, more researches used zonally averaged data, since a great part of the variation of ozone can be explained by physical processes with zonal symmetries.
	Liu \textit{et al.} \cite{liuQuasibiennialSemiannualOscillation2011} found that the dynamic transport was the principal factor controlling the QBO pattern of ozone and the QBO signals of ozone originated in the middle stratosphere and propagated downward along with the w-star anomalies over the equator.
	Xie \textit{et al.} \cite{xieImpactsTwoTypes2014} quantified the impacts on global ozone variation from two types of El Ni\~{n}o, i.e., EP-ENSO and CP-ENSO.
	For spatially distributed data, decomposition of total column ozone in the tropics revealed the spatial and temporal patterns of ozone variation \cite{campTemporalSpatialPatterns2003}, which showed strong relevance with  El Ni\~{n}o and QBO.
	There are vertical and horizontal channels in ozone transport at the same time, which can not be well described.
	Therefore, it is necessary to develop a new method to analyze the ozone over three dimensions (with longitude, latitude, and altitude).
	
	In this study, we introduce a theoretical framework, based on statistical physics and called eigen microstate method, to study the variation of ozone.
	Statistical physics approaches have been successfully applied to understand the complex Earth system \cite{fanStatisticalPhysicsApproaches2020}, especially the climate network method \cite{dijkstraNetworksClimate2019,agarwalQuantifyingRolesSingle2018,yamasakiClimateNetworksGlobe2008,dongesComplexNetworksClimate2009,ekhtiariDisentanglingMultiscaleEffects2019,agarwalNetworkbasedIdentificationCharacterization2019,agarwalOptimalDesignHydrometric2020,gaoNetworksFormedInterdependent2012,luImpactsNinoEvents2020}.
	For example, it improved the prediction of ENSO \cite{ludescherImprovedNinoForecasting2013,mengForecastingMagnitudeOnset2018}, and revealed the influence of Rossby waves in carbon dioxide ($\text{CO}_\text{2}$) concentration \cite{yingRossbyWavesDetection2020,yingClimateNetworkApproach2021,yingClimateNetworksSuggest2020}, air pollution \cite{zhangSignificantImpactRossby2019} and extreme-rainfall teleconnections \cite{boersComplexNetworksReveal2019}.
	Eigen microstate method \cite{sunEigenMicrostatesTheir2021} is a newly developed method, and can effectively analyze the emerging phenomena and the dynamic evolution of complex systems in non-equilibrium, such as the complex climate systems.
	It can be also used to identify the phase transition and universality class of  complex systems \cite{Hu2019}.
	The variation of ozone influenced by physical processes such as QBO or El Ni\~{n}o is expected to show collective behaviors, which can be regarded as an emerging phenomenon in complex climate systems.
	Therefore the eigen microstate method has the potential to better understand the ozone system, and to trace the influences from atmosphere circulations.
	
	To dispel the possible doubts about the eigen microstate method and EOF (empirical orthogonal function), we feel it is necessary to emphasize the differences here. The novelty and advantages of our methods in comparison to EOF are: 1) eigen microstate method originating from statistical physics provides a clearer physical insight of the climate systems, EOF is more like a data processing method; 2) microstates can be defined flexibly, different microstates can reflect different aspects of the complex systems; 3) the ability of eigen microstate method in describing phase transition.
	
	Here we apply the eigen microstate method to analyze the modes and their evolutions of global ozone concentration at different pressure levels. 
	This allows us to identify the changes of effects from different physical processes on ozone variance with geopotential heights. 
	Besides, the eigenvalues varying with geopotential heights can also reflect the change of atmosphere structure.
	In \cref{matherials}, we introduce the data and methods;
	In \Cref{result} we report the results and discussions. 
	Finally, we give a conclusion in \cref{discussion}.
	
	\section{Materials and Methods}\label{matherials}
	
	\subsection{Data}
	
	In this study, we have used the daily data of ozone mass mixing ratio (OMMR) provided by ERA5 (fifth generation ECMWF reanalysis for the global climate and weather) \cite{hersbachh.bellb.berrisfordp.biavatig.horanyia.munozsabaterj.nicolasj.peubeyc.radur.rozumi.schepersd.simmonsa.socic.deed.thepautj-n.ERA5HourlyData2018}. 
	In particular, we have chosen the OMMR data at 0 o'clock for every day from 1979/01/01 to 2020/06/30, containing $T = 15188$ days. Since we are interested at the intra-annual and inter-annual variance of OMMR, the change in one day can be neglected.
	The data set is 3-dimension, i.e., longitude, latitude, and altitude.
	It contains 37 pressure layers from \SI{1}{\hecto\pascal}, on the top of the stratosphere, to \SI{1000}{\hecto\pascal} near the surface of earth.
	The horizontal resolution is $2.5^{\circ}\times 2.5^{\circ}$, resulting in $ N = 72 \times 144 = 10368$ grids.
	
	\subsection{Eigen microstate method}
	
	The eigen microstate method \cite{sunEigenMicrostatesTheir2021} has shown its great ability to deal with the collective behaviors in complex systems.
	Here, we introduce how to implement this method into the ozone system.
	For a given complex system composed of $N$ agents, every grid is regarded as one agent, we can construct the microstates from OMMR data.
	The OMMR of grid $i$ at pressure layer $k$ and time $t$ is denoted as $O_i^k(t)$.
	The average ratio $\bar {O_i^k}$ during the considered time at grid $i$ is calculated as $ \bar {O_i^k } = \frac{1}{T}\sum_{t = 1}^T O_i^k(t)$.
	At time $t$, the OMMR fluctuation $ \delta O_i^k(t)$ at grid $i$ can be defined
	as $\delta O_i^k(t) = O_i^k(t) - \bar {O_i^k} $.
	The standard derivations of OMMR for the polar and equatorial regions are quite different, thus we construct the microstate $\bm{S}^k(t) = [s_1^k(t), s_2^k(t), ..., s_N^k(t)]$ at time $t$ and pressure layer $k$ by the normalized fluctuation of OMMR,
	\begin{equation}
		s_i^k(t) = \frac{ \delta O_i^k(t)}{ \sqrt{ \langle [\delta O_i^k(t) ]^2  \rangle }}\,,
	\end{equation}\label{equ:s} 
	where the denominator $\sqrt{ \langle [\delta O_i^k(t) ]^2  \rangle }$ stands for the standard derivation of the OMMR time series at grid $i$ and pressure layer $k$.
	
	Based on the above defined microstates $\bm{S}^k(t)$, we can compose a statistic ensemble of this complex system for every pressure layer $k$.
	The ensemble can be presented as an $N \times T$ matrix $\bm{A}^k$, with element $A_{it}^k$ defined as
	
	\begin{equation}
		A_{it}^k = \frac{s_i^k(t)}{\sqrt{C^k_0}}\,,
	\end{equation}
	where the normalization term $C^k_0 = \sum_{i = 1}^N \sum_{t = 1}^T s_i^k(t)^2$.
	Using singular value decomposition (SVD), the ensemble matrix $\bm{A}^k$ can be factorized as
	
	\begin{equation}\label{eq:svd}
		\bm{A}^k = \bm{U}^k \cdot \bm{\Lambda}^k\cdot (\bm{V}^k)^T =  \sum_{j = 1} \lambda_j^k \bm{u}_j^k \otimes (\bm{v}_j^k)^T\,,
	\end{equation}
	where $\bm{U}^k = [ \bm{u}_1^k, \bm{u}_2^k, ..., \bm{u}_N^k ] $ and $\bm{V}^k = [ \bm{v}_1^k, \bm{v}_2^k, ..., \bm{v}_T^k ] $ are both orthogonal matrices; $\lambda_j^k$ is the diagonal element of diagonal matrix $\bm{\Lambda}^k$ and is numbered in order $\lambda_1^k \ge \lambda_2^k\ge...\ge\lambda_T^k$.
	
	The correlation $C_{ij}^k$ between grids $i$ and $j$ is defined as $C_{ij}^k = \bm{S}_i^k \cdot (\bm{S}_j^k)^T $, where $\bm{S}_i^k$ and $\bm{S}_j^k$ are normalized OMMR fluctuation time series at grids $i$ and $j$ respectively.
	Then the correlation matrix can be written as $\bm{C}^k = C_0^k \bm{A}^k \cdot(\bm{A}^k)^T $.
	After taking into \cref{eq:svd}, we can get
	
	\begin{equation}
		\dfrac{\bm{C}^k}{C_0^k} \cdot \bm{u}_j^k = (\lambda_j^k)^2 \bm{u}_j^k\,.
	\end{equation}
	Here the trace of matrix $\bm{C}^k / C_0^k$ is equal 1, then $\sum_{i = 1}^T (\lambda_i^k)^2 = 1 $. $\bm{u}_j^k$ stands for the $j\text{th}$ eigen microstate showing collective behaviors and the value of $( \lambda_j^k)^2$ stands for the weight of eigen microstate $j$.
	Manipulated with the OMMR fluctuation time series, the evolution of the eigen microstate $j$ can be calculated as
	
	\begin{equation}
		\bm{S}^{e,k}_j= \sum_{i = 1}^N \bm{S}_i^k u_{ij}^k = \lambda_j^k \bm{v}_j^k\,.
	\end{equation}
	Thus, $\bm{v}_j^k$ is proportional to the evolution of the \textit{j}th eigen microstate $\bm{S}^{e,k}_j$, then $\bm{v}_j^k$ stands for the evolution of the \textit{j}th eigen microstate.
	
	\subsection{"LOTUS\_regression" model} \label{sec: lrm}
	
	To quantify the relation between the eigen microstates and different physical processes, we take advantage of the LOTUS\_regression \cite{SPARCIO3CGAW2020}.
	The LOTUS\_regression is a multi-linear regression model designed to derive long-term trends of ozone by LOTUS (Long-term Ozone Trends and their Uncertainties in the Stratosphere) group.
	This LOTUS\_regression model takes the natural processes causing ozone  varying  as proxies in the linear regression, and gets the trend of ozone after eliminating the influence the proxies.
	The selected proxies contain two orthogonal components of the QBO, the solar \SI{10.7}{\centi\meter} flux, ENSO without any lag applied, and the AOD (Aerosol Optical Depth).
	This regression can be fulfilled easily through the software supplied by LOTUS group.
	The result of the regression is to obtain the coefficients A-J that correspond to the following equation:
	
	\begin{align}
		p(t) = & A\cdot qboA(t) + B\cdot qboB(t) + C\cdot ENSO(t) + D\cdot AOD(t) + E \cdot Solar(t) + Trop(t)\notag\\
		& + F\cdot linear_{pre}(t) + G\cdot linear_{post}(t) +H \cdot C_1(t) + I\cdot C_2(t) +J \cdot C_3(t) +\epsilon(t)\,.
	\end{align}
	
	The data of considered processes can be found on the open database (see \cref{tab:proxies}), except that $\epsilon(t)$ is the remaining error and $linear_{pre}(t)$, $linear_{post}(t)$, $C_1(t)$, $C_2(t)$ and $C_3(t)$ are the linear terms \cite{SPARCIO3CGAW2020}.
	$qboA(t)$ and $qboB(t)$ are the first two orthogonal components from monthly mean zonal wind data at levels 70, 50, 40, 30, 20, 15, and \SI{10}{\hecto\pascal} with principal component analysis.
	$ENSO(t)$ is the MEI (the bi-monthly multivariate ENSO index) calculated by  five different variables (sea level pressure (SLP), sea surface temperature (SST), zonal and meridional components of the surface wind, and outgoing longwave radiation (OLR)) over the tropical Pacific basin\cite{wolterMeasuringStrengthENSO1998, zhangProbabilisticMultivariateENSO2019}.
	$AOD(t)$ and $Solar(t)$ stand for the mean aerosol optical depth at \SI{550}{\nano\meter} and the solar \SI{10.7}{\centi\meter} flux respectively. 
	$Trop(t)$ represents the time series of the height of tropopause \cite{kalnayNCEPNCAR40Year1996}.
	
	All the time series of proxies, when applied to regression, has been normalized, thus the standard derivation is 1 and the mean value is 0.
	Then the coefficient before the proxy is proportional how much it contributes to $p(t)$.

\begin{table}[!ht]
	\begin{adjustwidth}{-1.5in}{0in} 
		\centering
		\caption{Data sources of proxies included in the LOTUS\_regression}\label{tab:proxies}
		\begin{tabular}{cc}
		\hline
		\textbf{variable}	& \textbf{source}\\
		\hline
		$qboA$, $qboB$		& http://www.geo.fu-berlin.de/met/ag/strat/produkte/qbo/ \\
		$ENSO$	& http://www.esrl.noaa.gov/psd/enso/mei\\
		$AOD$     & https://data.giss.nasa.gov/modelforce/strataer/ \\
		$Solar$   & http://www.spaceweather.ca/data-donnee/sol\_flux/sx-5-mavg-eng.php\\
		$Trop$   & https://www.esrl.noaa.gov/psd/data/gridded/data.ncep.reanalysis.tropopause.html\\ \hline
		\end{tabular}
	\end{adjustwidth}
\end{table}

	\section{Results and Discussions}\label{result}
	
	We calculate the eigen microstates and their evolution of OMMR both for raw data and deseasonalized data.
	The deseasonalized time series of OMMR $O^\prime(y, d)$ can be obtained by removing the yearly mean value, thus $O'(y,d) = O(y,d) - \frac{1}{N(d)}\sum_y O(y,d)$. Here $y$ stands for the year, $d$ stands for the day and $N(d)$ is the number of years containing day $d$.
	In the following, notations with prime at the top right stand for variables obtained from deseasonalized data. For example, $\bm{u}_1$ is the first eigen microstate calculated from raw data while  $\bm{u}'_1$ from deseasonalized data.
	
	\subsection{Intra-annual oscillation from raw data}
	
\begin{figure}
		\includegraphics[width=8.5cm]{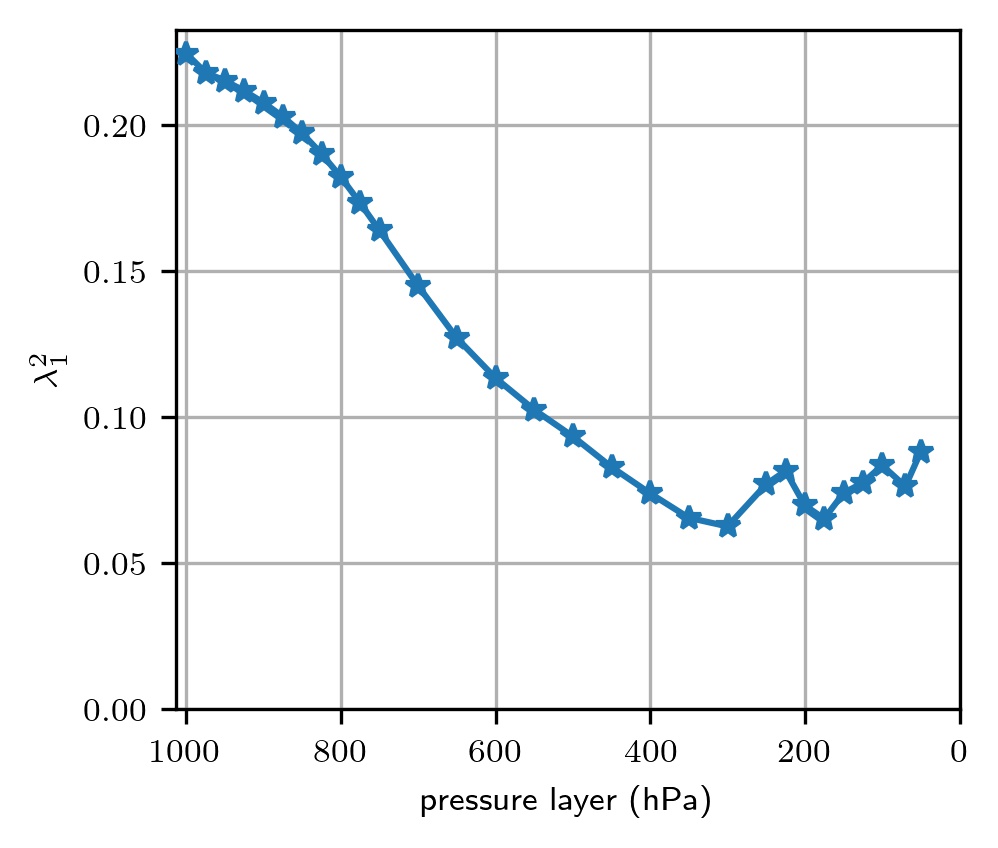}
		\caption{The weight $\lambda_1^2$ of the first eigen microstate $\bm{u}_1$ calculated from the raw data varies with pressure layers.}\label{fig:weight1}
	\end{figure}

	For the raw data, we find that the evolution $\bm{v}_1$ of the first eigen microstate has a strong one-year cycle for almost all pressure layers.
	The annual cycle is related to the seasonal effect caused by the Earth's revolutions around the Sun.
	\Cref{fig:weight1} presents the curve of the relation between the weight $\lambda_1^2$ of the first eigen microstate $\bm{u}_1$ and the geopotential height. It shows that the weight $\lambda_1^2$ decreases from 47\% at \SI{1000}{\hecto\pascal} to nearly 10\% at \SI{300}{\hecto\pascal} then increases to 40\% at \SI{50}{\hecto\pascal} again.
	It means that the seasonal effect along OMMR is higher at 1000 and \SI{50}{\hecto\pascal}, lower at around \SI{300}{\hecto\pascal}.
	The curve of weight $\lambda_1^2$ is similar to atmospheric temperature profile \cite{rodgersRetrievalAtmosphericTemperature1976}, with the tipping point at \SI{300}{\hecto\pascal}.
	And it is directly opposite to the trend identified in vertical structure of tropic temperature whose tipping point is also \SI{300}{\hecto\pascal} \cite{trenberthVerticalStructureTemperature2006}.
	
	\begin{figure}
		\includegraphics[width=8.5 cm]{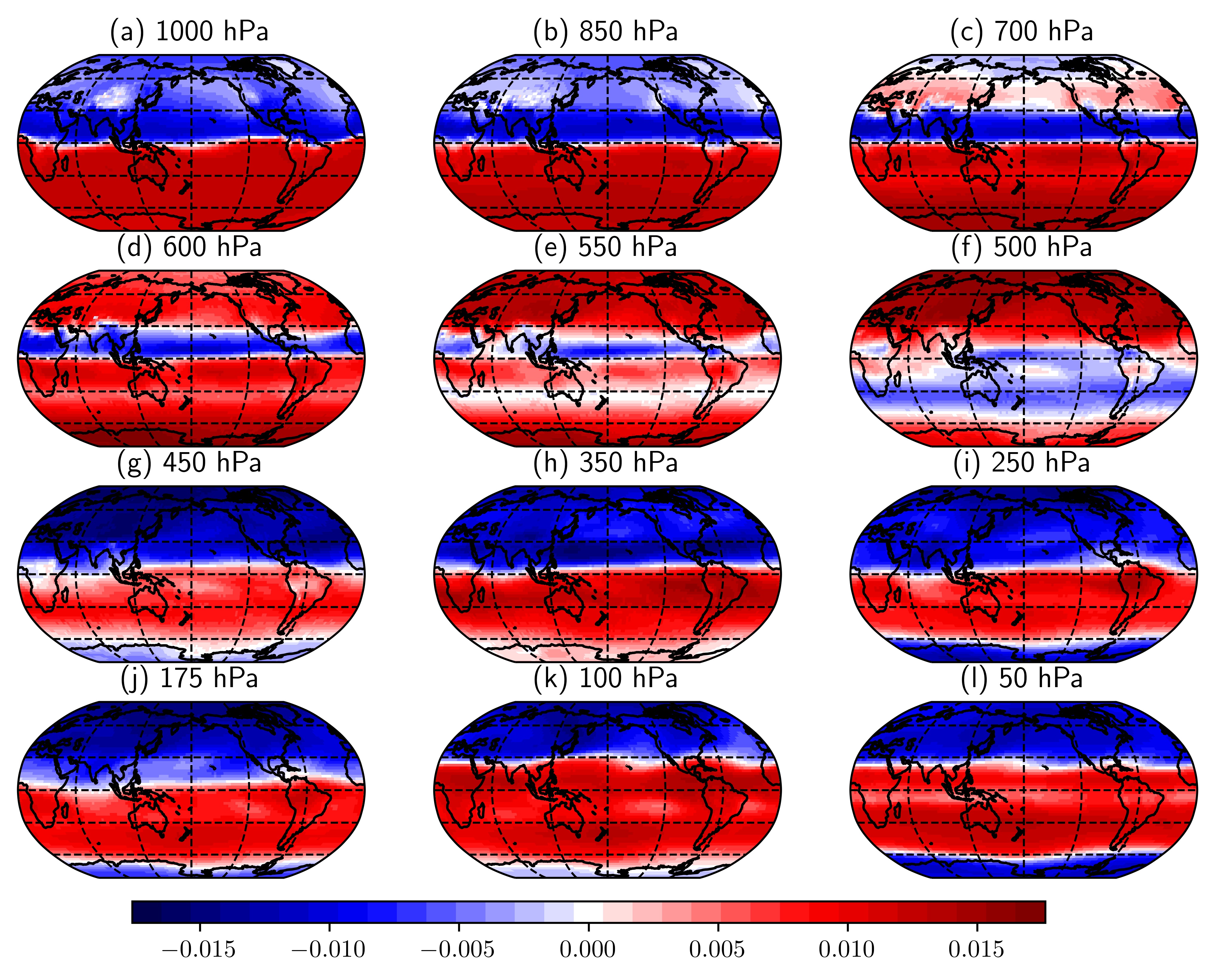}
		\caption{Spatial distributions of the first eigen microstate $\bm{u}_1$ at
			\SI{1000}{\hecto\pascal}, \SI{850}{\hecto\pascal}, \SI{700}{\hecto\pascal}, \SI{600}{\hecto\pascal}, \SI{550}{\hecto\pascal}, \SI{500}{\hecto\pascal}, \SI{450}{\hecto\pascal}, \SI{350}{\hecto\pascal}, \SI{250}{\hecto\pascal}, \SI{175}{\hecto\pascal}, \SI{100}{\hecto\pascal} and \SI{50}{\hecto\pascal} are depicted at panel (a), (b), (c), (d), (e), (f), (g), (h), (i), (j), (k) and (l), respectively.
			The colorbar with zero in the middle is fixed for all subfigures, and the color corresponds to the value of the first eigen microstates.
		}\label{fig:mode1}
	\end{figure}
	
	\begin{figure}
		\includegraphics[width=8.5 cm]{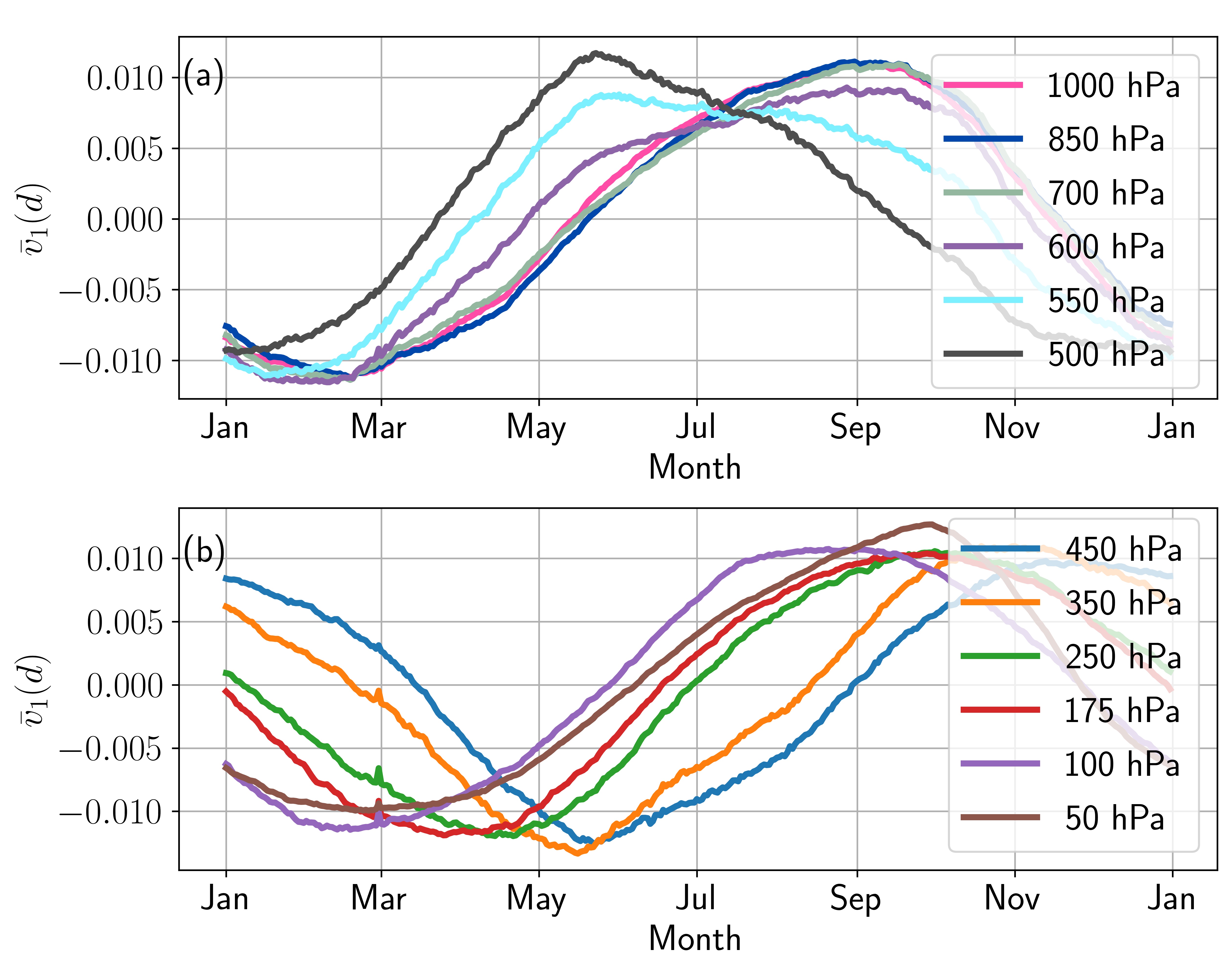}
		\caption{The intra-annual oscillation $\bar{v}_1(d)$ dominating in the evolution $\bm{v}_1$ of first eigen microstate. (a) Panel contains geopotential heights from \SI{1000}{\hecto\pascal} to \SI{500}{\hecto\pascal}; (b) panel contains geopotential heights from \SI{450}{\hecto\pascal} to \SI{50}{\hecto\pascal}.}\label{fig:state1}
	\end{figure}
	
	To further describe how the season effect changes in OMMR  with pressure levels, we show the spatial distribution of the first microstate  $\bm{u}_1$ (see \cref{fig:mode1}) and the intra-annual oscillation $\bar{v}_1(d)$ dominating the evolution of the first microstate (see \cref{fig:state1}) at \SI{1000}{\hecto\pascal}, \SI{850}{\hecto\pascal}, \SI{700}{\hecto\pascal}, \SI{600}{\hecto\pascal}, \SI{550}{\hecto\pascal}, \SI{500}{\hecto\pascal}, \SI{450}{\hecto\pascal}, \SI{350}{\hecto\pascal}, \SI{250}{\hecto\pascal}, \SI{175}{\hecto\pascal}, \SI{100}{\hecto\pascal} and \SI{50}{\hecto\pascal}, respectively.
	The intra-annual oscillation $\bar{v}_1(d)$ of the evolution of the first microstate can be obtained by $\bar{v}_1(d) = \frac{1}{N(d)}\sum_y v_1(y,d)$.
	
	\Cref{fig:mode1}a, b, c show the spatial distribution of the first eigen microstates near the Earth's surface. We find that they are characterized by two major clusters: one is located in the northern hemisphere (colored as blue), and the other one in the southern hemisphere (colored as red). 
	Besides, the intra-annual oscillation of the evolution of these first eigen microstates are nearly the same, see \cref{fig:state1}a, shown as a sinusoidal wave: it gets the smallest around February and largest in September.
	It is opposite to the temperature, for high temperature restrains the production of ozone.
	When we consider the eigen microstates at the geopotential height up to \SI{550}{\hecto\pascal}, although the frequency of the first state is still annual cycle, the globe except for a small region near the equator is colored as red (see \cref{fig:mode1}d, e), which means that the globe behaves in the same way except for the region near the equator in the opposite way.
	In addition, the intra-annual oscillations at \SI{600}{\hecto\pascal}, \SI{550}{\hecto\pascal} shown in \cref{fig:state1}a are distorted from standard sinusoidal oscillation. 
	This could be caused by some atmospheric circulations with the period of one year strongly coupled with the seasonal cycle around \SI{550}{\hecto\pascal}, such as Hadley cells \cite{diazHadleyCirculationPresent2004}.
	Finally, the spatial distribution of the first eigen microstates, located at geopotential heights from the top of the troposphere to the middle of the stratosphere (see \cref{fig:mode1}g-l), show a similar pattern to the one near the surface. But there are still two differences that the dividing line around the equator moves towards to the north, and finally lies at $30^\circ N$; the south pole shows different colors as the altitude increases.
	Moreover, the intra-annual oscillations of these eigen microstates' evolution still maintain the shape of a sinusoidal wave, while the phase decreases with height. \Cref{fig:state1}b shows that the time reaching the minimum is advanced from June to early February as the altitude increases from \SI{450}{\hecto\pascal} to \SI{50}{\hecto\pascal}.
	
	\subsection{The relation between eigen microstates and atmospheric circulations}
	
	To better analyze the mechanism of the emergence of collective behaviors in OMMR, we remove the seasonal effect in the following analysis.
	For a complex systems without localization of microstate, the weight of an eigen microstate approaches zero in the thermodynamic limit \cite{sunEigenMicrostatesTheir2021}.
	But for the system of ozone, the variation is affected by both natural and anthropogenic variation. 
	The system should behave a condensation of eigen microstates, which could be caused by several atmosphere circulations at the same time.
	For example, Camp \textit{et al.} showed the conflation of decadal and QBO signal in the first two EOFs (empirical orthogonal function).
	Since it is hard to estimate the influence degree of a certain circulation from frequency analysis.
	Here, we use the "LOTUS\_regression" model to identify the eigen microstates dominated by trend, tropopause pressure, ENSO and QBO, respectively.
	
	\subsubsection{Trend}
	
\begin{figure}
		\includegraphics[width=14 cm]{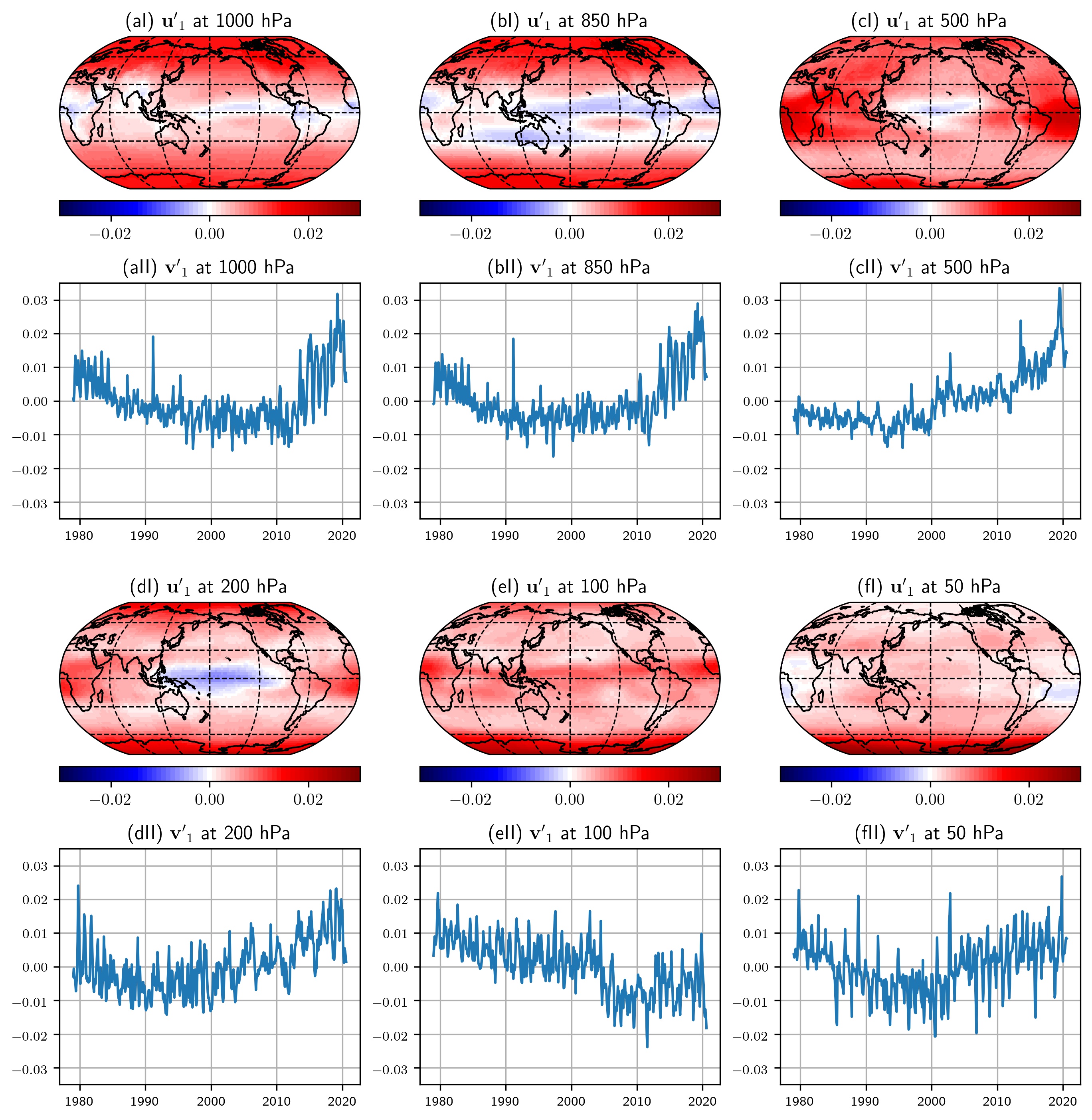}
		\caption{The first eigen microstate $\bm{u}^\prime_1$ calculated from deseasonalized data are shown at panel index as \Romannum{1}, and their time evolution $\bm{v}^\prime_1$ below the eigen microstate with panel index as \Romannum{2}. Six typical pressure layers are considered, i.e., 1000, 850, 500, 200, 100, \SI{50}{\hecto\pascal}. The eigen microstates and their time evolution are depicted at penal (a), (b), (c), (d), (e) and (f), respectively.}\label{fig:mode1ds}
	\end{figure}

	After deseasonalization, the first eigen microstates can usually extract the trend of OMMR for different pressure layers.
	As shown in \cref{fig:mode1ds}, the globe is nearly colored as red except that the eigen microstates at \SI{850}{\hecto\pascal} and \SI{200}{\hecto\pascal} where showing a slight blue around the equator.
	For the pressure layers near the surface (see \cref{fig:mode1ds}a\Romannum{2}, b\Romannum{2}), the ozone concentration is decreasing from 1979 to 1990, keeps steady from 1990 to 2010, and then increases rapidly after 2010.
	When considering \SI{50}{\hecto\pascal} at the middle of the ozone layer (see \cref{fig:mode1ds}f\Romannum{2}), we  identify the effect from Montreal Protocol: OMMR is decreasing until the mid-1990s and it began to recover after the 2000s.
	The oscillation is strong, and the time deeply going down could be corresponded to the period when the ozone hole was appearing.
	The trends of ozone between the surface and ozone layer can be influenced strongly by the vertical transport (see \cref{fig:mode1ds}c\Romannum{2}, d\Romannum{2}, e\Romannum{2}).
	
	\subsubsection{Tropopause Pressure}   
	
	Affected by the temperature, the ozone mixing ratio profile in the lower stratosphere shifts up and down with the tropopause \cite{steinbrechtCorrelationsTropopauseHeight1998}. The positive relation between the total ozone column and tropopause height has already been confirmed by Camp \cite{campTemporalSpatialPatterns2003}.
	
\begin{figure}
		\includegraphics[width=14 cm]{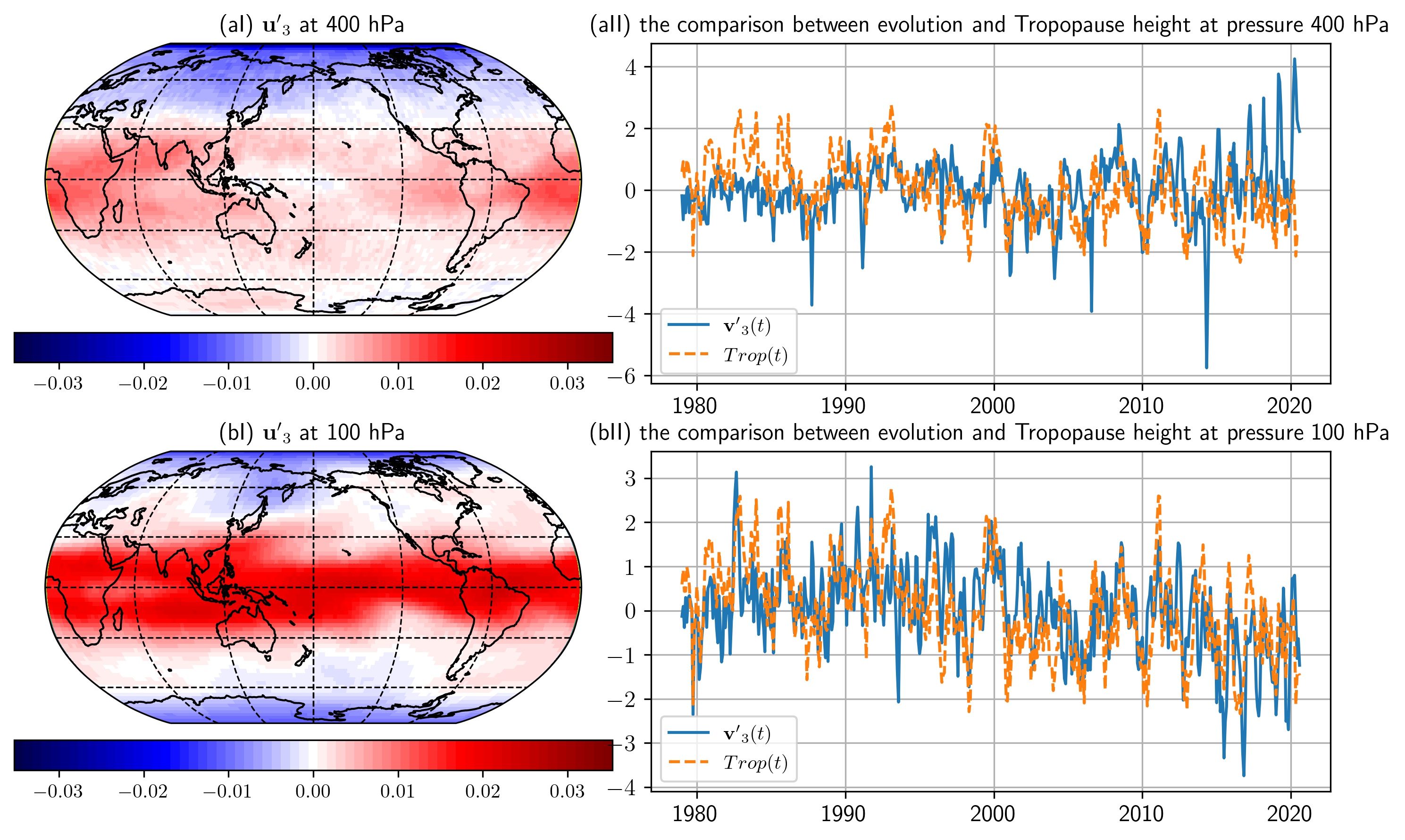}
		\caption{The third eigen microstates $\bm{u}_3$ and their evolution with time $\bm{v}_3(t)$ at pressure 400 and \SI{100}{\hecto\pascal} are shown at panel (a) and (b), respectively. The monthly average time series of evolution with time is compared to $Trop(t)$ introduced at \cref{sec: lrm}.}\label{fig:mode3ds}
	\end{figure}

	In this study, we find that the high correlation between the tropopause height and the top several eigen microstates only exists at the layers from around \SI{400}{\hecto\pascal} to \SI{50}{\hecto\pascal}, and is the strongest at \SI{100}{\hecto\pascal}. 
	The thrid eigen microstates and their evolution at pressure layer of 400, \SI{100}{\hecto\pascal} are shown in \cref{fig:mode3ds}. 
	For the third eigen microstates at 400, \SI{100}{\hecto\pascal} and their evolution, LOTUS\_regression shows that the dominating proxy is tropopause pressure.
	
	The Pearson correlation coefficient (denoted as $r$ in the following) between the evolution time series $\bm{v}_3$ and $Trop(t)$ is 0.57 at \SI{100}{\hecto\pascal}, much stronger than 0.29 at \SI{400}{\hecto\pascal}.
	\Cref{fig:mode3ds}b\Romannum{1} shows that the region highly positively correlated to tropopause pressure is mainly located at low latitude region, i.e., between $30^\circ S$ and $30^\circ N$.
	
	\subsubsection{ENSO}
	
\begin{figure}
		\includegraphics[width=14 cm]{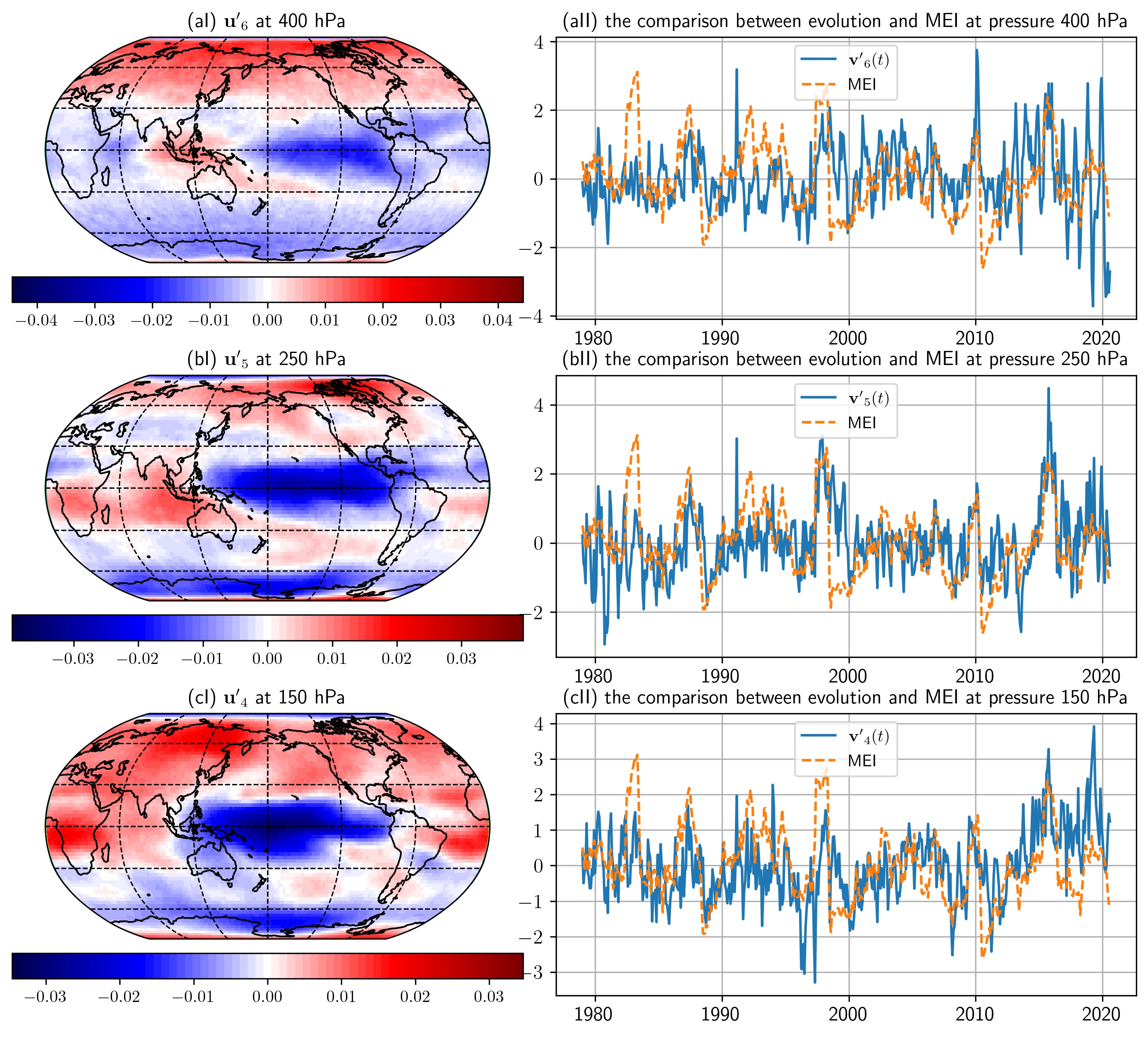}
		\caption{
			The sixth eigen microstate at \SI{400}{\hecto\pascal}, fifth eigen microstate at \SI{250}{\hecto\pascal} and fourth eigen microstate at \SI{150}{\hecto\pascal} are depicted at penal (a\Romannum{1}), (b\Romannum{1}) and (c\Romannum{1}), respectively.
			The monthly average time series of their evolution are plotted at right, and are compared with MEI introduced at \cref{sec: lrm}.}\label{fig:enso}
	\end{figure}

	\Cref{fig:enso} shows three typical eigen microstates: the sixth eigen microstate at \SI{400}{\hecto\pascal}, the fifth eigen microstate at \SI{250}{\hecto\pascal}, the fourth eigen microstate at \SI{150}{\hecto\pascal}, and their corresponding evolution with time.
	The ENSO pattern does not appear at the same eigen microstate for different geopotential heights.
	We find that these eigen microstates (see \cref{fig:enso}a\Romannum{1}, b\Romannum{1}, c\Romannum{1}) show a characteristic bimodal structure around the central Pacific. 
	And the bimodal structure is moving toward the west as the geopotential height rising.
	Besides, \cref{fig:enso}a\Romannum{2}, b\Romannum{2}, c\Romannum{2} show the evolution of eigen microstates at 400, 250, and \SI{150}{\hecto\pascal} compared to the MEI.
	The Pearson correlation coefficients calculated by $\bm{v}'_6$ at \SI{400}{\hecto\pascal}, $\bm{v}'_5$ at \SI{250}{\hecto\pascal} and $\bm{v}'_4$ at \SI{150}{\hecto\pascal} are 0.24, 0.45 and 0.47, respectively.
	Because we choose the values of the above eigen microstates at ENSO region as negative, the variation of ozone is negatively related to MEI.
	It can be seen that the pressure layer strongly influenced by ENSO lies between the upper layer of troposphere and the lower layer of stratosphere. The ozone at \SI{400}{\hecto\pascal} is slightly impacted with negligible $r$ as 0.24.
	
	ENSO can affect tropical upwelling \cite{wmoworldmeteorologicalorganizationScientificAssessmentOzone2018}, which in turn leads to fluctuations in temperature and ozone in the tropical lower stratosphere.
	Our layered analysis indicates that eigen microstates dominated by ENSO mainly appear at the 4th, 5th, 6th eigen microstates between \SI{400}{\hecto\pascal} and \SI{150}{\hecto\pascal}.	
	The influence from ENSO on the distribution of OMMR from around \SI{400}{\hecto\pascal} to around \SI{150}{\hecto\pascal} enhances with the geopotential heights. 
	
	\subsubsection{QBO}
	
\begin{figure}
		\includegraphics[width=14 cm]{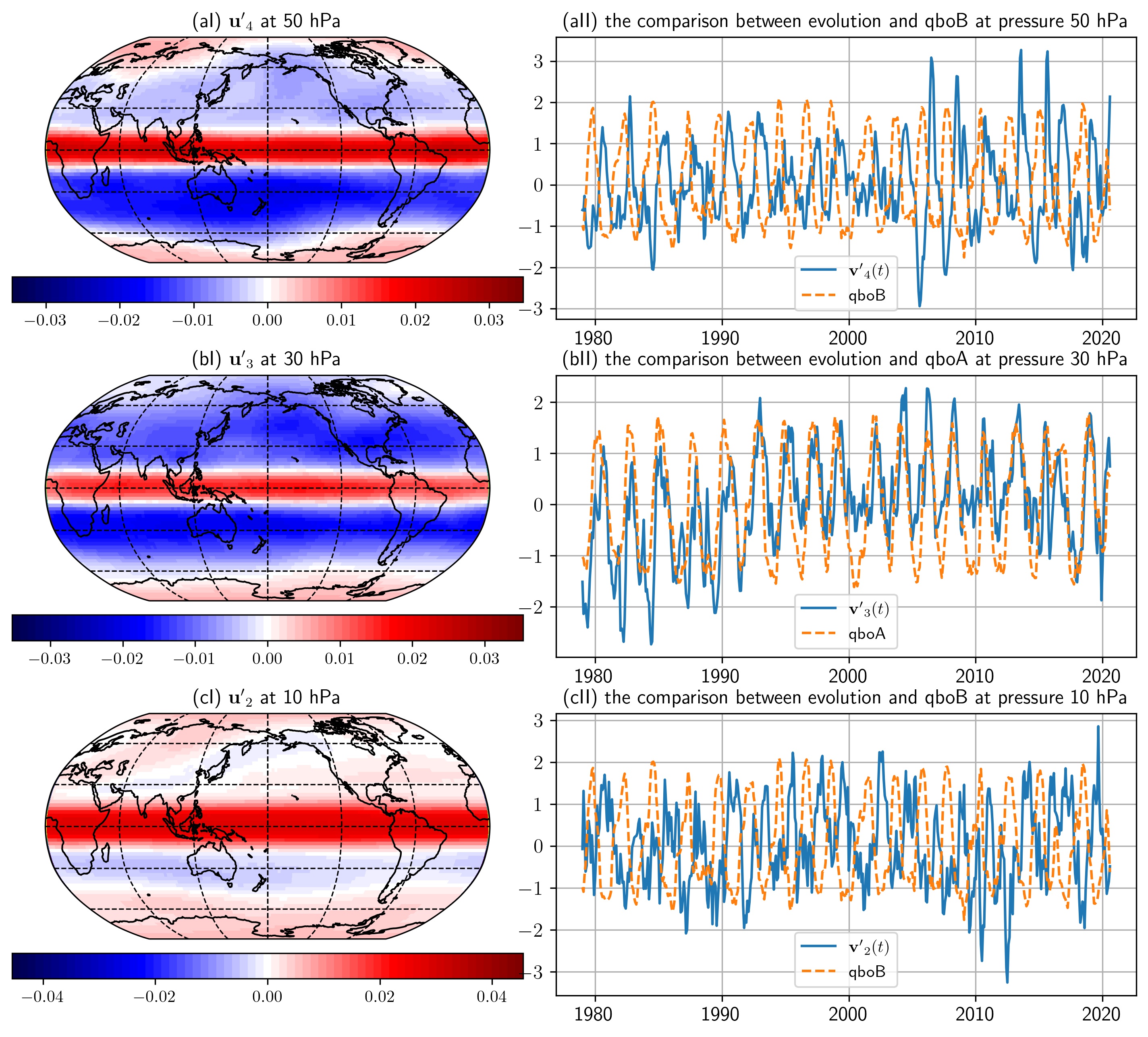}
		\caption{The fourth eigen microstate at \SI{50}{\hecto\pascal}, third eigen microstate at \SI{30}{\hecto\pascal} and second eigen microstate at \SI{10}{\hecto\pascal} are depicted at penal (a\Romannum{1}), (b\Romannum{1}) and (c\Romannum{1}), respectively.	The monthly average time series of their evolution are plotted at right, and are compared with $qboA$ or $qboB$ introduced at \cref{sec: lrm}.}\label{fig:qbods}
	\end{figure}

	\Cref{fig:qbods} shows three typical eigen microstates: the fourth eigen microstate at \SI{50}{\hecto\pascal}, the third eigen microstate at \SI{30}{\hecto\pascal}, the second eigen microstate at \SI{10}{\hecto\pascal}, and their evolution with time.
	All the eigen microstates reveal that the area within $15^\circ S-15^\circ N$ is highly correlated. In addition, \cref{fig:qbods}a\Romannum{1} shows one "blue" cluster in the middle latitude region of southern hemisphere, \cref{fig:qbods}b\Romannum{1} shows two "blue" clusters: one at southern hemisphere and the other at the middle latitude region of southern hemisphere.
	The evolution of the fourth eigen microstate at \SI{50}{\hecto\pascal} and the second eigen microstate at \SI{10}{\hecto\pascal} is highly negatively correlated with $qboB$ with $r$ as -0.6 and -0.59, respectively.
	The evolution of the third eigen microstate at \SI{30}{\hecto\pascal} is highly positively correlated with $qboA$ with $r$ as 0.67.
	
	The ozone distribution at stratosphere could be highly influenced by QBO \cite{naoeFutureChangesOzone2017, wmoworldmeteorologicalorganizationScientificAssessmentOzone2018}.
	The QBO signal in tropical ozone consists of a primary maximum in amplitude at around \SI{7}{\hecto\pascal}, a secondary maximum near 20-\SI{30}{\hecto\pascal}, and a minimum near \SI{15}{\hecto\pascal}. 
	This is self-consistent with our results. 
	
	\section{Conclusions}\label{discussion}
	
	In this article, we have calculated the eigen microstates and their evolution with time for both raw data and deseasonalized data of OMMR at different geopotential heights. For raw data,
	1) the curve of weights of the first eigen microstate to geopotential heights is similar to temperature profile. This can reveal the change of atmosphere structure;
	2) the time evolution $\bm{v}_1$ of the first eigen microstate indicates a strong annual cycle and the phase of the derived intra-annual oscillation from $\bm{v}_1$ is changing with the geopotential heights.
	For deseasonalized data, however, we used the "LOTUS\_regression" model to identify the relation between eigen microstates and the atmosphere circulations, including trend, tropopause pressure, ENSO and QBO. We find that 
	\romannum{1}) the first eigen microstate $\bm{u'}_1$ usually captures the overall trends of OMMR at different pressure layers;
	\romannum{2}) tropopause pressure impacts eigen microstates at pressure layer between around \SI{400}{\hecto\pascal} and \SI{100}{\hecto\pascal}, and the influence is strongest at \SI{100}{\hecto\pascal}; 
	\romannum{3}) the typical ENSO pattern at Pacific is quite clear for eigen microstates between pressure layer of \SI{400}{\hecto\pascal} and \SI{150}{\hecto\pascal}, and is moving toward west with the increase of geopotential heights, and the influence of ENSO on eigen microstate enhances as the geopotential height increases;
	\romannum{4}) the QBO signal is also identified well for the pressure layers from \SI{50}{\hecto\pascal} to \SI{10}{\hecto\pascal}.
	
	Our framework presented here not only facilitates the deep understanding of ozone but can potentially also be used as a measure for other complex Earth systems. For example, the distribution of carbon dioxide in space and ocean currents could be potential and fruitful research fields. Besides, there is also a chance to identify the tipping points of AMOC (Atlantic meridional overturning circulation), Arctic summer sea-ice, Amazon rainforest and more with our framework.
	
	
	\section*{Acknowledgments}
	We wish to acknowledge Qing Yao for reading the manuscript and providing some useful comments. This work was support by Key Research Program of Frontier Sciences, Chinese Academy of Sciences (Grant No. QYZD-SSW-SYS019).
	
	\nolinenumbers
	
	\bibliography{emso3}

\begin{thebibliography}{10}

\bibitem{agarwalNetworkbasedIdentificationCharacterization2019}
A.~Agarwal, L.~Caesar, N.~Marwan, R.~Maheswaran, B.~Merz, and J.~Kurths.
\newblock Network-based identification and characterization of teleconnections
  on different scales.
\newblock {\em Scientific Reports}, 9(1):8808, 2019.

\bibitem{agarwalQuantifyingRolesSingle2018}
A.~Agarwal, N.~Marwan, R.~Maheswaran, B.~Merz, and J.~Kurths.
\newblock Quantifying the roles of single stations within homogeneous regions
  using complex network analysis.
\newblock {\em Journal of Hydrology}, 563:802--810, 2018.

\bibitem{agarwalOptimalDesignHydrometric2020}
A.~Agarwal, N.~Marwan, R.~Maheswaran, U.~Ozturk, J.~Kurths, and B.~Merz.
\newblock Optimal design of hydrometric station networks based on complex
  network analysis.
\newblock {\em Hydrology and Earth System Sciences}, 24(5):2235--2251, 2020.

\bibitem{boersComplexNetworksReveal2019}
N.~Boers, B.~Goswami, A.~Rheinwalt, B.~Bookhagen, B.~Hoskins, and J.~Kurths.
\newblock Complex networks reveal global pattern of extreme-rainfall
  teleconnections.
\newblock {\em Nature}, 566(7744):373--377, 2019.

\bibitem{brauerAmbientAirPollution2016}
M.~Brauer, G.~Freedman, J.~Frostad, A.~{van Donkelaar}, R.~V. Martin,
  F.~Dentener, R.~van Dingenen, K.~Estep, H.~Amini, J.~S. Apte,
  K.~Balakrishnan, L.~Barregard, D.~Broday, V.~Feigin, S.~Ghosh, P.~K. Hopke,
  L.~D. Knibbs, Y.~Kokubo, Y.~Liu, S.~Ma, L.~Morawska, J.~L.~T. Sangrador,
  G.~Shaddick, H.~R. Anderson, T.~Vos, M.~H. Forouzanfar, R.~T. Burnett, and
  A.~Cohen.
\newblock Ambient {{Air Pollution Exposure Estimation}} for the {{Global
  Burden}} of {{Disease}} 2013.
\newblock {\em Environmental Science \& Technology}, 50(1):79--88, 2016.

\bibitem{bronnimannVariabilityTotalOzone2000}
S.~Br{\"o}nnimann, J.~Luterbacher, C.~Schmutz, H.~Wanner, and J.~Staehelin.
\newblock Variability of total ozone at {{Arosa}}, {{Switzerland}}, since 1931
  related to atmospheric circulation indices.
\newblock {\em Geophysical Research Letters}, 27(15):2213--2216, 2000.

\bibitem{campTemporalSpatialPatterns2003}
C.~D. Camp, M.~S. Roulston, and Y.~L. Yung.
\newblock Temporal and spatial patterns of the interannual variability of total
  ozone in the tropics.
\newblock {\em Journal of Geophysical Research: Atmospheres}, 108(D20), 2003.

\bibitem{diazHadleyCirculationPresent2004}
H.~F. Diaz and R.~S. Bradley.
\newblock The {{Hadley Circulation}}: {{Present}}, {{Past}}, and {{Future}}.
\newblock In H.~F. Diaz and R.~S. Bradley, editors, {\em The {{Hadley
  Circulation}}: {{Present}}, {{Past}} and {{Future}}}, Advances in {{Global
  Change Research}}, pages 1--5. {Springer Netherlands}, {Dordrecht}, 2004.

\bibitem{dijkstraNetworksClimate2019}
H.~A. Dijkstra, E.~{Hern{\'a}ndez-Garc{\'i}a}, C.~Masoller, and M.~Barreiro.
\newblock {\em Networks in {{Climate}}}.
\newblock {Cambridge University Press}, {Cambridge}, 2019.

\bibitem{dongesComplexNetworksClimate2009}
J.~F. Donges, Y.~Zou, N.~Marwan, and J.~Kurths.
\newblock Complex networks in climate dynamics.
\newblock {\em The European Physical Journal Special Topics}, 174(1):157--179,
  2009.

\bibitem{ekhtiariDisentanglingMultiscaleEffects2019}
N.~Ekhtiari, A.~Agarwal, N.~Marwan, and R.~V. Donner.
\newblock Disentangling the multi-scale effects of sea-surface temperatures on
  global precipitation: {{A}} coupled networks approach.
\newblock {\em Chaos: An Interdisciplinary Journal of Nonlinear Science},
  29(6):063116, 2019.

\bibitem{fanStatisticalPhysicsApproaches2020}
J.~Fan, J.~Meng, J.~Ludescher, X.~Chen, Y.~Ashkenazy, J.~Kurths, S.~Havlin, and
  H.~J. Schellnhuber.
\newblock Statistical physics approaches to the complex {{Earth}} system.
\newblock {\em Physics Reports}, 2020.

\bibitem{fengGroundlevelO3Pollution2015}
Z.~Feng, E.~Hu, X.~Wang, L.~Jiang, and X.~Liu.
\newblock Ground-level {{O3}} pollution and its impacts on food crops in
  {{China}}: {{A}} review.
\newblock {\em Environmental Pollution}, 199:42--48, 2015.

\bibitem{gaoNetworksFormedInterdependent2012}
J.~Gao, S.~V. Buldyrev, H.~E. Stanley, and S.~Havlin.
\newblock Networks formed from interdependent networks.
\newblock {\em Nature Physics}, 8(1):40--48, 2012.

\bibitem{hermanGlobalAverageOzone1991}
J.~R. Herman, R.~McPeters, R.~Stolarski, D.~Larko, and R.~Hudson.
\newblock Global average ozone change from {{November}} 1978 to {{May}} 1990.
\newblock {\em Journal of Geophysical Research: Atmospheres},
  96(D9):17297--17305, 1991.

\bibitem{hersbachh.bellb.berrisfordp.biavatig.horanyia.munozsabaterj.nicolasj.peubeyc.radur.rozumi.schepersd.simmonsa.socic.deed.thepautj-n.ERA5HourlyData2018}
{Hersbach, H., Bell, B., Berrisford, P., Biavati, G., Hor\'anyi, A., Mu\~noz
  Sabater, J., Nicolas, J., Peubey, C., Radu, R., Rozum, I., Schepers, D.,
  Simmons, A., Soci, C., Dee, D., Th\'epaut, J-N.}
\newblock {{ERA5}} hourly data on pressure levels from 1979 to present.
  {{Copernicus Climate Change Service}} ({{C3S}}) {{Climate Data Store}}
  ({{CDS}}).
\newblock 2018.

\bibitem{Hu2019}
G.~K. Hu, T.~Liu, M.~X. Liu, W.~Chen, and X.~S. Chen.
\newblock Condensation of eigen microstate in statistical ensemble and phase
  transition.
\newblock {\em Science China: Physics, Mechanics and Astronomy}, 62(9), 2019.

\bibitem{SPARCIO3CGAW2020}
{I. Petropavlovskikh}, {S. Godin-Beekmann}, {D. Hubert}, {R. Damadeo}, {B.
  Hassler}, and {V. Sofieva}.
\newblock {{SPARC}}/{{IO3C}}/{{GAW Report}} on {{Long}}-term {{Ozone Trends}}
  and {{Uncertainties}} in the {{Stratosphere}}.
\newblock {\em SPARC Report No. 9, GAW Report No. 241, WCRP-17/2018}, 2019.

\bibitem{jockelAtmosphericChemistryGeneral2006}
P.~J{\"o}ckel, H.~Tost, A.~Pozzer, C.~Br{\"u}hl, J.~Buchholz, L.~Ganzeveld,
  P.~Hoor, A.~Kerkweg, M.~G. Lawrence, R.~Sander, B.~Steil, G.~Stiller,
  M.~Tanarhte, D.~Taraborrelli, J.~{van Aardenne}, and J.~Lelieveld.
\newblock The atmospheric chemistry general circulation model
  {{ECHAM5}}/{{MESSy1}}: Consistent simulation of ozone from the surface to the
  mesosphere.
\newblock {\em Atmospheric Chemistry and Physics}, 6(12):5067--5104, 2006.

\bibitem{kalnayNCEPNCAR40Year1996}
E.~Kalnay, M.~Kanamitsu, R.~Kistler, W.~Collins, D.~Deaven, L.~Gandin,
  M.~Iredell, S.~Saha, G.~White, J.~Woollen, Y.~Zhu, M.~Chelliah, W.~Ebisuzaki,
  W.~Higgins, J.~Janowiak, K.~C. Mo, C.~Ropelewski, J.~Wang, A.~Leetmaa,
  R.~Reynolds, R.~Jenne, and D.~Joseph.
\newblock The {{NCEP}}/{{NCAR}} 40-{{Year Reanalysis Project}}.
\newblock {\em Bulletin of the American Meteorological Society},
  77(3):437--472, 1996.

\bibitem{karimiConcentrationsHealthEffects2019}
A.~Karimi, M.~Shirmardi, M.~Hadei, Y.~T. Birgani, A.~Neisi, A.~Takdastan, and
  G.~Goudarzi.
\newblock Concentrations and health effects of short- and long-term exposure to
  {{PM2}}.5, {{NO2}}, and {{O3}} in ambient air of {{Ahvaz}} city, {{Iran}}
  (2014\textendash 2017).
\newblock {\em Ecotoxicology and Environmental Safety}, 180:542--548, 2019.

\bibitem{liuQuasibiennialSemiannualOscillation2011}
Y.~Liu, C.~Lu, Y.~Wang, and E.~Kyr{\"o}l{\"a}.
\newblock The quasi-biennial and semi-annual oscillation features of tropical
  {{O}} 3 , {{NO}} 2 , and {{NO}} 3 revealed by {{GOMOS}} satellite
  observations for 2002\textendash 2008.
\newblock {\em Chinese Science Bulletin}, 56(18):1921--1929, 2011.

\bibitem{luImpactsNinoEvents2020}
Z.~Lu, N.~Yuan, L.~Chen, and Z.~Gong.
\newblock On the {{Impacts}} of {{El Ni\~no Events}}: {{A New Monitoring
  Approach Using Complex Network Analysis}}.
\newblock {\em Geophysical Research Letters}, 47(6):e2019GL086533, 2020.

\bibitem{ludescherImprovedNinoForecasting2013}
J.~Ludescher, A.~Gozolchiani, M.~I. Bogachev, A.~Bunde, S.~Havlin, and H.~J.
  Schellnhuber.
\newblock Improved {{El Ni\~no}} forecasting by cooperativity detection.
\newblock {\em Proceedings of the National Academy of Sciences},
  110(29):11742--11745, 2013.

\bibitem{mengForecastingMagnitudeOnset2018}
J.~Meng, J.~Fan, Y.~Ashkenazy, A.~Bunde, and S.~Havlin.
\newblock Forecasting the magnitude and onset of {{El Ni\~no}} based on climate
  network.
\newblock {\em New Journal of Physics}, 20(4):043036, 2018.

\bibitem{naoeFutureChangesOzone2017}
H.~Naoe, M.~Deushi, K.~Yoshida, and K.~Shibata.
\newblock Future {{Changes}} in the {{Ozone Quasi}}-{{Biennial Oscillation}}
  with {{Increasing GHGs}} and {{Ozone Recovery}} in {{CCMI Simulations}}.
\newblock {\em Journal of Climate}, 30(17):6977--6997, 2017.

\bibitem{nowackLargeOzonecirculationFeedback2015}
P.~J. Nowack, N.~L. Abraham, A.~C. Maycock, P.~Braesicke, J.~M. Gregory, M.~M.
  Joshi, A.~Osprey, and J.~A. Pyle.
\newblock A large ozone-circulation feedback and its implications for global
  warming assessments.
\newblock {\em Nature Climate Change}, 5(1):41--45, 2015.

\bibitem{oltmansSurfaceOzoneDistributions1986}
S.~J. Oltmans and W.~D. Komhyr.
\newblock Surface ozone distributions and variations from 1973\textendash 1984:
  {{Measurements}} at the {{NOAA Geophysical Monitoring}} for {{Climatic Change
  Baseline Observatories}}.
\newblock {\em Journal of Geophysical Research: Atmospheres},
  91(D4):5229--5236, 1986.

\bibitem{polvaniStratosphericOzoneDepletion2011}
L.~M. Polvani, D.~W. Waugh, G.~J.~P. Correa, and S.-W. Son.
\newblock Stratospheric {{Ozone Depletion}}: {{The Main Driver}} of
  {{Twentieth}}-{{Century Atmospheric Circulation Changes}} in the {{Southern
  Hemisphere}}.
\newblock {\em Journal of Climate}, 24(3):795--812, 2011.

\bibitem{rodgersRetrievalAtmosphericTemperature1976}
C.~D. Rodgers.
\newblock Retrieval of atmospheric temperature and composition from remote
  measurements of thermal radiation.
\newblock {\em Reviews of Geophysics}, 14(4):609--624, 1976.

\bibitem{staehelinOzoneTrendsReview2001}
J.~Staehelin, N.~R.~P. Harris, C.~Appenzeller, and J.~Eberhard.
\newblock Ozone trends: {{A}} review.
\newblock {\em Reviews of Geophysics}, 39(2):231--290, 2001.

\bibitem{steinbrechtCorrelationsTropopauseHeight1998}
W.~Steinbrecht, H.~Claude, U.~K{\"o}hler, and K.~P. Hoinka.
\newblock Correlations between tropopause height and total ozone:
  {{Implications}} for long-term changes.
\newblock {\em Journal of Geophysical Research: Atmospheres},
  103(D15):19183--19192, 1998.

\bibitem{sunEigenMicrostatesTheir2021}
Y.~Sun, G.~Hu, Y.~Zhang, B.~Lu, Z.~Lu, J.~Fan, X.~Li, Q.~Deng, and X.~Chen.
\newblock Eigen microstates and their evolutions in complex systems.
\newblock {\em Communications in Theoretical Physics}, 73(6):065603, 2021.

\bibitem{trenberthVerticalStructureTemperature2006}
K.~E. Trenberth and L.~Smith.
\newblock The {{Vertical Structure}} of {{Temperature}} in the {{Tropics}}:
  {{Different Flavors}} of {{El Ni\~no}}.
\newblock {\em Journal of Climate}, 19(19):4956--4973, 2006.

\bibitem{wmoworldmeteorologicalorganizationScientificAssessmentOzone2018}
{WMO (World Meteorological Organization)}.
\newblock Scientific {{Assessment}} of {{Ozone Depletion}}: 2018.
\newblock Global {{Ozone Research}} and {{Monitoring
  Project}}\textendash{{Report}} No. 58, {Geneva, Switzerland}, 2018.

\bibitem{wolterMeasuringStrengthENSO1998}
K.~Wolter and M.~S. Timlin.
\newblock Measuring the strength of {{ENSO}} events: {{How}} does 1997/98 rank?
\newblock {\em Weather}, 53(9):315--324, 1998.

\bibitem{xieImpactsTwoTypes2014}
F.~Xie, J.~Li, W.~Tian, J.~Zhang, and J.~Shu.
\newblock The impacts of two types of {{El Ni\~no}} on global ozone variations
  in the last three decades.
\newblock {\em Advances in Atmospheric Sciences}, 31(5):1113--1126, 2014.

\bibitem{xuEstimatingImpactGround2021}
M.~Xu, Q.~Yao, D.~Chen, M.~Li, R.~Li, B.~Gao, B.~Zhao, and Z.~Chen.
\newblock Estimating the impact of ground ozone concentrations on crop yields
  across {{China}} from 2014 to 2018: {{A}} multi-model comparison.
\newblock {\em Environmental Pollution}, 283:117099, 2021.

\bibitem{yamasakiClimateNetworksGlobe2008}
K.~Yamasaki, A.~Gozolchiani, and S.~Havlin.
\newblock Climate {{Networks}} around the {{Globe}} are {{Significantly
  Affected}} by {{El Ni}}\textbackslash\textasciitilde no.
\newblock {\em Physical Review Letters}, 100(22):228501, 2008.

\bibitem{yingClimateNetworkApproach2021}
N.~Ying, W.~Wang, J.~Fan, D.~Zhou, Z.~Han, Q.~Chen, Q.~Ye, and Z.~Xue.
\newblock Climate network approach reveals the modes of {{CO2}} concentration
  to surface air temperature.
\newblock {\em Chaos: An Interdisciplinary Journal of Nonlinear Science},
  31(3):031104, 2021.

\bibitem{yingClimateNetworksSuggest2020}
N.~Ying, D.~Zhou, Z.~Han, Q.~Chen, Q.~Ye, Z.~Xue, and W.~Wang.
\newblock Climate networks suggest {{Rossby}}-waves\textendash related {{CO2}}
  concentrations to surface air temperature.
\newblock {\em EPL (Europhysics Letters)}, 132(1):19001, 2020.

\bibitem{yingRossbyWavesDetection2020}
N.~Ying, D.~Zhou, Z.~G. Han, Q.~H. Chen, Q.~Ye, and Z.~G. Xue.
\newblock Rossby {{Waves Detection}} in the {{CO2}} and {{Temperature
  Multilayer Climate Network}}.
\newblock {\em Geophysical Research Letters}, 47(2):e2019GL086507, 2020.

\bibitem{zhangProbabilisticMultivariateENSO2019}
T.~Zhang, A.~Hoell, J.~Perlwitz, J.~Eischeid, D.~Murray, M.~Hoerling, and T.~M.
  Hamill.
\newblock Towards {{Probabilistic Multivariate ENSO Monitoring}}.
\newblock {\em Geophysical Research Letters}, 46(17-18):10532--10540, 2019.

\bibitem{zhangSignificantImpactRossby2019}
Y.~Zhang, J.~Fan, X.~Chen, Y.~Ashkenazy, and S.~Havlin.
\newblock Significant {{Impact}} of {{Rossby Waves}} on {{Air Pollution
  Detected}} by {{Network Analysis}}.
\newblock {\em Geophysical Research Letters}, 46(21):12476--12485, 2019.

\end{thebibliography}
	
	\bibliographystyle{abbrv}
	
\end{document}